\pdfoutput=1
\documentclass[cits]{PoS}

\usepackage{amsmath}

\newcommand{\gev}{\,\operatorname{GeV}}

\newcommand{\ms}{\mskip 1.5mu}

\renewcommand{\vec}[1]{\textbf{#1}}

\title{\raggedleft{\small DESY 13-113}  \\[2em] \raggedright
  From form factors to generalized parton distributions}

\ShortTitle{From form factors to generalized parton distributions}

\author{\speaker{Markus DIEHL}\\
  Deutsches Elektronen-Synchroton DESY, 22603 Hamburg, Germany \\
  E-mail: \email{markus.diehl@desy.de}}

\abstract{I present an extraction of generalized parton distributions
  from selected data on the electromagnetic nucleon form factors.  The
  extracted distributions can in particular be used to quantify the
  contribution to the proton spin from the total angular momentum
  carried by valence quarks, as well as their transverse spatial
  distribution inside the proton.}

\FullConference{XXI International Workshop on Deep-Inelastic Scattering
  and Related Subjects-DIS2013,\\
  22-26 April 2013\\
  Marseilles,France}

\begin{document}

\section{Introduction}

Generalized parton distributions (GPDs) contain unique information about
the structure of the proton.  Most prominently, they quantify the
transverse spatial distributions of partons in correlation with their
longitudinal momentum \cite{Burkardt:2002hr,Diehl:2002he}.  Specific GPDs
provide access to the angular momentum carried by partons, as becomes
manifest from Ji's sum rule \cite{Ji:1998pc} and from spin-orbit
correlations such as the change of the spatial parton distribution induced
by transverse proton polarization \cite{Burkardt:2002hr}.

Sum rules connect the GPDs for unpolarized quarks with the electromagnetic
nucleon form factors according to
\begin{align}
  \label{sum-rules}
F_1^q(t) &= \int_{0}^1 dx\; H_v^q(x,t) \,,
&
F_2^q(t) &= \int_{0}^1 dx\; E_v^q(x,t) \,,
\intertext{where we introduced the valence (quark minus antiquark) combinations}
  \label{valence-GPDs}
H_v^q(x,t) &= H^q(x,0,t) + H^q(-x,0,t) \,,
&
E_v^q(x,t) &= E^q(x,0,t) + E^q(-x,0,t)
\end{align}
of the conventional GPDs $H^q(x,\xi,t)$ and $E^q(x,\xi,t)$ defined in
\cite{Ji:1998pc}.  The Dirac and Pauli form factors of the proton are
given by charge weighted sums $F_{i}^p(t) = \sum_{q} e_q\, F^q_{i}(t)$
with $i=1,2$, and corresponding expressions for the neutron form factors
are obtained using isospin symmetry.  A Fourier transform to transverse
position space turns $H_v^q$ into the probability density for finding a
quark with flavor $q$, longitudinal momentum fraction $x$ and transverse
position $\vec{b}$ in a proton, minus the corresponding probability
density for antiquarks.

This contribution presents the work in ref.~\cite{Diehl:2013xca}, where
GPDs are determined from nucleon form factor data via the sum rules
\eqref{sum-rules}.  In several ways, such a study is complementary to GPD
extractions from hard exclusive processes like deeply virtual Compton
scattering or meson production~\cite {Sabatie:DIS2013}.
$(i)$ Form factor data extend to large values of $-t$, much beyond what
can ever be reached in hard exclusive processes.
$(ii)$ The integrals in \eqref{sum-rules} are sensitive to a wide range of
$x$, from about $10^{-3}$ to $0.6$ and higher in our analysis.
$(iii)$ The sum rules \eqref{sum-rules} can be written in terms of GPDs at
zero skewness $\xi$, which after Fourier transform admit a probability
interpretation and obey much simpler positivity bounds than their
counterparts with $\xi\neq 0$.
$(iv)$ On the downside, it is impossible to uniquely reconstruct the $x$
and $t$ dependence of GPDs from \eqref{sum-rules}, which leaves us with an
irreducible model dependence due to the functional forms assumed for
$H_v^q(x,t)$ and $E_v^q(x,t)$.


\section{Form factor data}

For our analysis we have made a selection from the available data on
electromagnetic form factors.  The magnetic proton from factor $G_M^p(t)$
offers the largest reach in $t$, up to $-t \simeq 31 \gev^2$.  With
experimental uncertainties of only a few $\%$ at small to intermediate
$t$, quantitative control of two-photon exchange in the elastic $ep$ cross
section is essential for a reliable determination of this quantity.  We
use the data extracted in the global analysis of \cite{Arrington:2007ux}
and checked that it is compatible with the results of other recent work.

\begin{figure}
\begin{center}
\includegraphics[width=0.49\textwidth,%
  viewport=20 0 345 250]{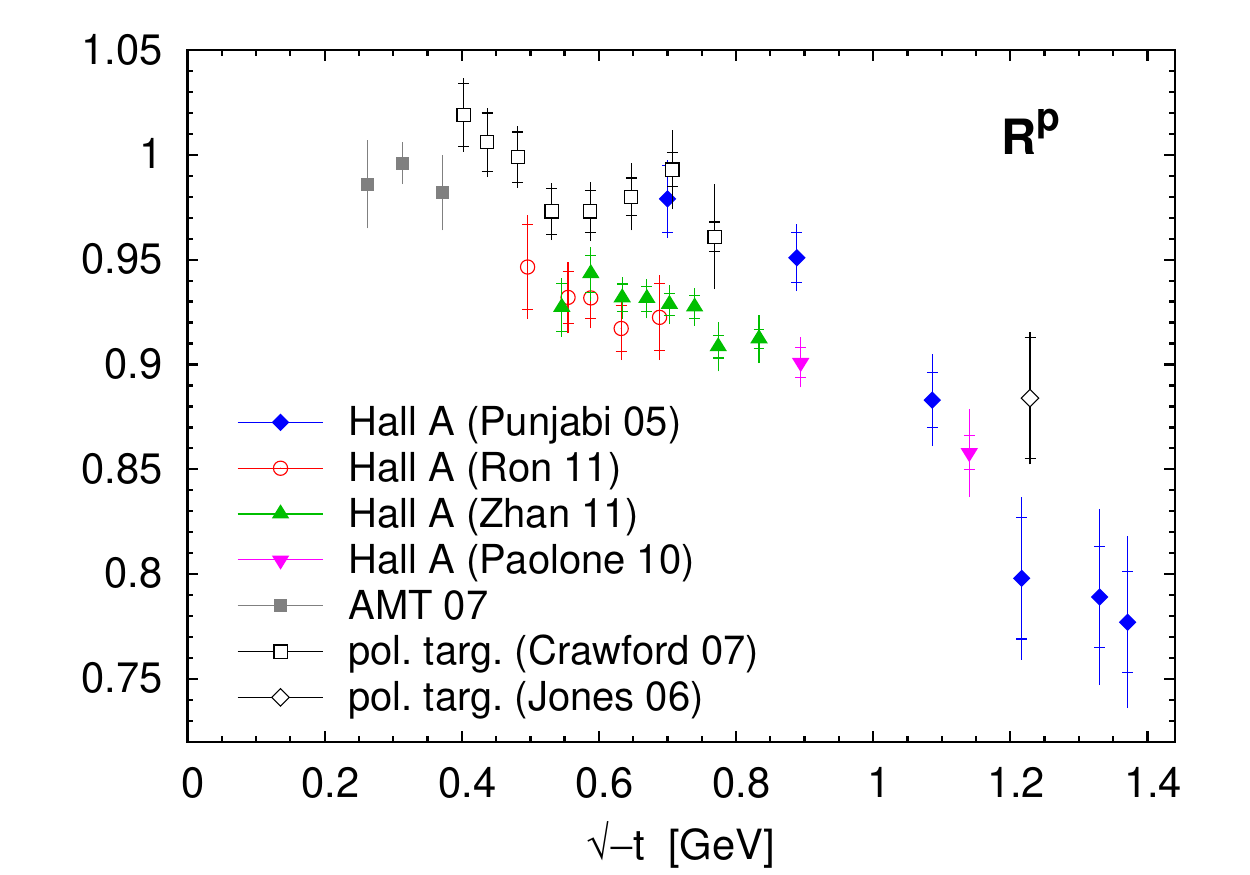}
\includegraphics[width=0.49\textwidth,%
  viewport=20 0 345 250]{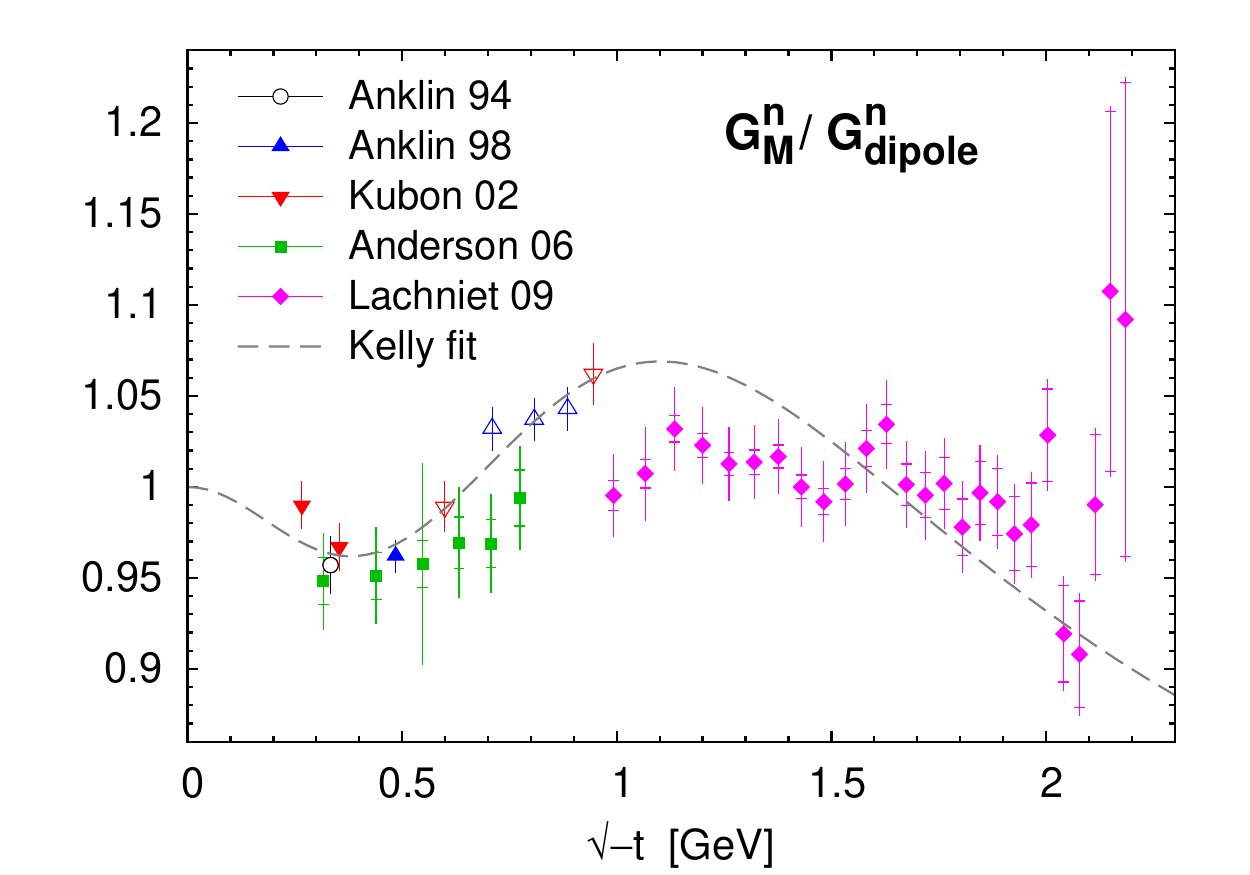}
\end{center}
\caption{\label{fig:data} Left: Selection of recent data on the scaled
  from factor ratio $R^p = \mu_p\ms G_E^p / G_M^p$.  Right: Data on
  $G_M^n$, divided by the conventional dipole parameterization.  Open
  symbols denote points that are omitted in our analysis (but clearly
  dominated a fit by Kelly in 2004, shown as a dashed line).  References
  for the data sets are given in \protect\cite{Diehl:2013xca}.}
\end{figure}

The scaled ratio $R^p = \mu_p\ms G_E^p / G_M^p$ of electric and magnetic
form factors can be extracted from elastic $ep$ scattering data where
either the polarization of the recoil proton is measured or the proton
target is polarized.  As is evident from figure \ref{fig:data}, recoil
polarization data from JLab Hall A published in 2010 and 2011 are clearly
inconsistent with older data from the same hall and with measurements on
polarized proton targets.  We omit the polarized target data from our fit,
while emphasizing that a clarification of the experimental situation is
urgent.

Let us now turn to the neutron form factor data.  For $G_M^n$ there is a
number of data points from the 1990s that are inconsistent with more
recent determinations (see figure \ref{fig:data}) and have been excluded
from our fits.  For the electric neutron form factor $G_E^n$ or the ratio
$R^n = \mu_n\ms G_E^n / G_M^n$ there is an overall consistent data set,
although only up to $-t \simeq 3.4 \gev^2$.  In our fits we also include
the very precise determination of the squared charge radius $r_{nE}^2$
from neutron scattering on shell electrons in nuclear targets
\cite{Beringer:1900zz}.

As a benchmark for our fit using GPDs, we have performed a simple global
fit to the form factor data we have selected.  We use the functional form
\begin{align}
\label{global-power-fit-fct}
F_i^{q}(t) &= F_i^{q}(0)\,
  (1 - a_{iq}\, t/p_{iq})^{-p_{iq}}_{\phantom{1}} \,
  (1 - b_{iq}\, t/q_{iq})^{-q_{iq}}_{\phantom{1}}
\end{align}
with $q=u,d$ and $i=1,2$, fixing 3 of the 16 parameters in this ansatz and
fitting the remaining 13.  We obtain a global $\chi^2 = 122.3$ for 178
data points and a uniformly good description of the data for the proton
and neutron form factors.

Furthermore, we have interpolated the data on the Sachs form factors and
their ratios to a common set of $t$ values, where we can then compute mean
values and errors for the form factors $F_{i}^{q}(t)$ and for any
combination of them (see figure~\ref{fig:Fud-ratios} below).  Limited by
the data on $R^n$ and $G_E^n$, this set covers a range up to $-t \simeq
3.4 \gev^2$.

The electromagnetic form factors receive contributions also from strange
quarks, i.e.\ from $F_1^s$ and $F_2^s$.  We estimate these in a model that
is consistent with determinations from lattice QCD and from parity
violation in polarized elastic $ep$ scattering (with data up to $-t \simeq
0.6 \gev^2$).  We find that the estimated size of $F_{1,2}^s(t)$ is
comparable to the current \emph{uncertainties} on $F_{1,2}^{u,d}(t)$, so
that our poor knowledge of the strangeness sector does not yet have an
impact on our determination of the GPDs for $u$ and $d$ quarks.


\section{Fit of GPDs}

To extract valence quark GPDs from the form factor data we use an ansatz
developed in our earlier work \cite{Diehl:2004cx} (are related analysis
can be found in \cite{Guidal:2004nd}).  We take an exponential form
\begin{align}
  \label{HE-ansatz}
H_v^q(x,t) &=   q_v(x)\, \exp\bigl[ t \ms f_q(x) \bigr] \,,
&
E_v^q(x,t) &= e_v^q(x)\, \exp\bigl[ t \ms g_q(x) \bigr] \,,
\end{align}
for the $t$ dependence, with $x$ dependent slopes
\begin{align}
  \label{profile-ansatz}
f_q(x) &= \alpha_q'\ms (1-x)^3 \log(1/x) + B_q\ms (1-x)^3 
          + A_q\ms x (1-x)^2 \,,
\nonumber \\
g_q(x) &= \alpha_q'\ms (1-x)^3 \log(1/x) + D_q\ms (1-x)^3 
          + C_q\ms x (1-x)^2 \,.
\end{align}
A Fourier transform turns $H_v^q$ into a Gaussian in transverse space,
where the average squared impact parameter $\langle \vec{b}^2 \rangle^q_x
= 4 f_q(x)$ vanishes like $(1-x)^2$ for $x\to 1$ with our ansatz.  This
ensures that the distance $\vec{b} /(1-x)$ between the struck quark and
the spectators remains on average finite in this limit, which is plausible
for a system subject to confinement \cite{Burkardt:2002hr}.  The $t=0$
limit of $H_v^q$ is given by the usual valence quark densities.  Although
these are relatively well known, there are important differences between
different PDF sets at small and at large $x$.  We take the ABM 11 NLO
densities \cite{Alekhin:2012ig} at scale $\mu= 2\gev$ as our default and
repeat our analysis with a number of other recent parameterizations (CT
10, GJR 08, HERAPDF 1.5, MSTW 2008, NNPDF 2.2) as a cross check.  We have
verified that the CJ 12 parameterization, which was presented at this
conference \cite{Accardi:DIS2013} and differs from other sets in the
limiting behavior of $d_v$ at $x\to 1$, gives results within the range
spanned by the other parameterizations in our fits.
The $t=0$ limit of $E_v^q$ is unknown, and we make the
ansatz
\begin{equation}
  \label{e-ansatz}
e_v^q(x) \,\propto\, x^{-\alpha_q} (1-x)^{\beta_q}\,
           \bigl(1 + \gamma_q \sqrt{x} \ms\bigr) \,,
\end{equation}
where the normalization is determined by the anomalous magnetic moments of
the proton and neutron.  The Fourier transform of $E_v^q$ describes the
shift in the transverse spatial distribution of valence quarks that is
induced by transverse proton polarization \cite{Burkardt:2002hr}.  As a
consequence, the size of $E_v^q$ relative to $H_v^q$ is limited by a
positivity bound \cite{Burkardt:2003ck}.  Imposing this bound in our fit
significantly constrains the allowed parameter space, in particular by
limiting the large-$x$ power $\beta_q$ in $e_v^q$ from below.  Our best
fit results are obtained by taking both $\beta_u$ and $\beta_d$ as small
as possible, and we typically find $\beta_u \sim 4.5$ to $5$ and $\beta_d
\sim 5$ to $6$ depending on the PDF set used for $H_v^q$.  The precise
data on $r_{nE}^2$ and $R^n$ favor a small flavor dependence with
$\alpha'{\!\!}_u > \alpha'{\!\!}_d$ for the shrinkage parameter in $H_v^q$
and $E_v^q$.  Our fitted values of $\alpha'{\!\!}_d$ and $\alpha'{\!\!}_u$
range from $0.68$ to $1.0 \gev^{-2}$, which is in line with expectations
from Regge phenomenology, where the small-$x$ behavior of $H_v^q$ and
$E_v^q$ is determined by the $\rho$ and $\omega$ trajectories.

\begin{figure}
\begin{center}
\includegraphics[width=0.46\textwidth]{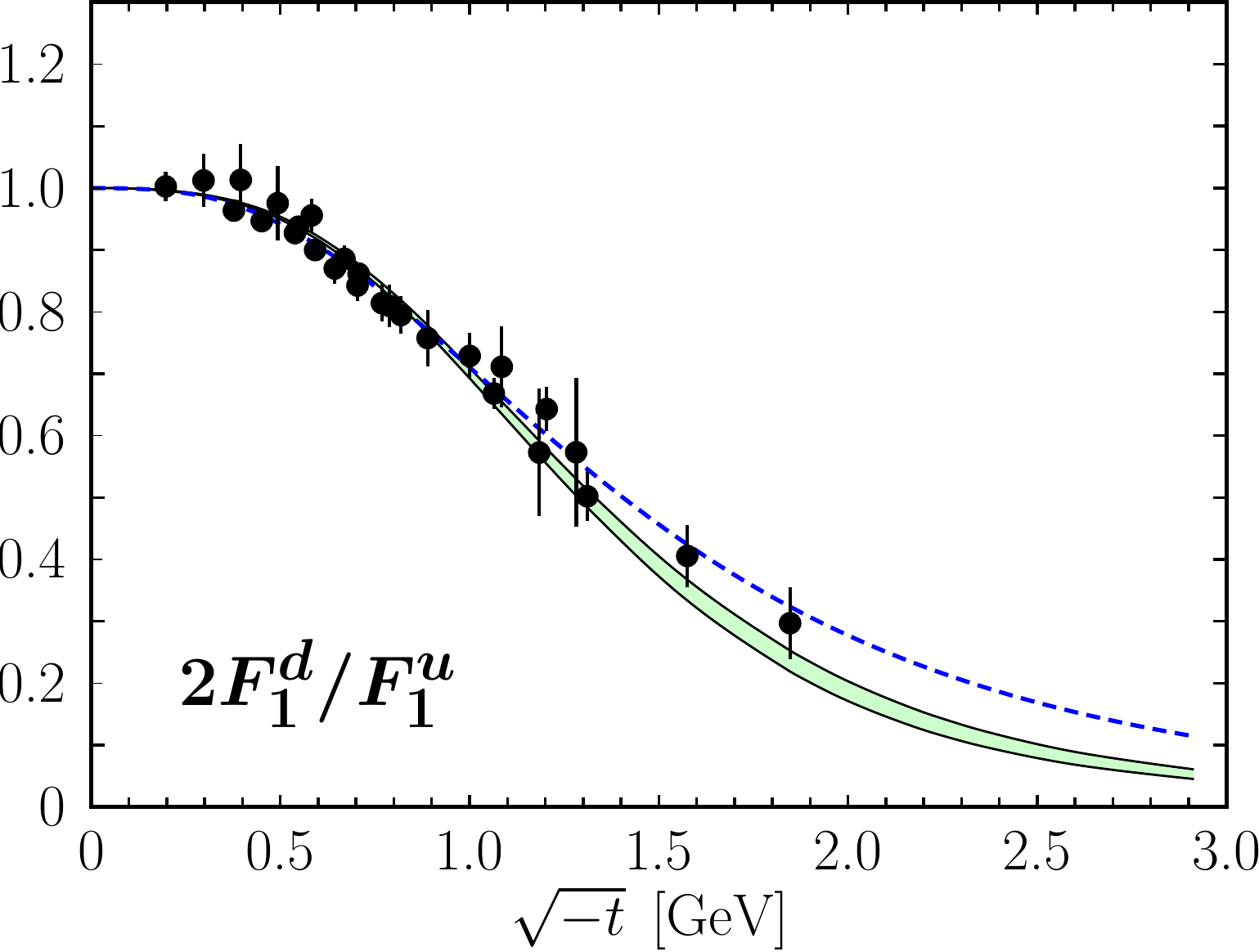}
\hspace{1em}
\includegraphics[width=0.46\textwidth]{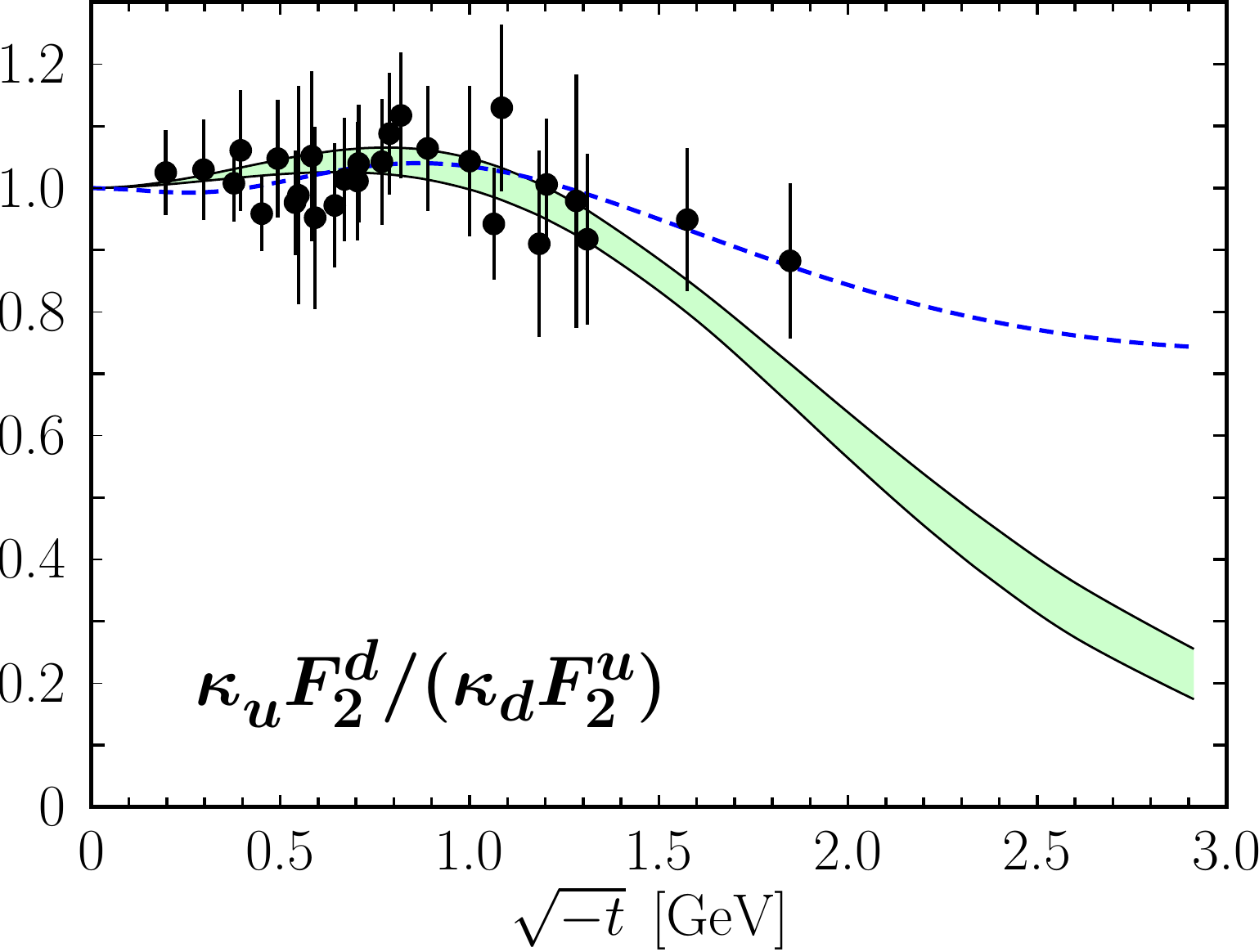}
\end{center}
\caption{\label{fig:Fud-ratios} Normalized ratios of the $d$ and $u$ quark
  contributions to the Dirac and Pauli form factors of the proton.  The
  band represents our default GPD fit and the dashed line the fit using
  \protect\eqref{global-power-fit-fct}.  The data points have been
  obtained by interpolation of data on Sachs form factors and their ratios
  to a set of common $t$ values.}
\end{figure}

We obtain a good overall fit of all form factor data, with a global
$\chi^2 = 221.2$ for $178$ data points if we use the ABM 11 PDFs.  There
are small discrepancies between fit and data for $G_M^p$ and $R^p$ below
$-t = 1\gev^2$, but only at the level of a few $\%$.  In
figure~\ref{fig:Fud-ratios} we show the ratios $F_1^d/F_1^u$ and
$F_2^d/F_2^u$, both normalized to $1$ at $t=0$.  As discussed in
\cite{Diehl:2004cx}, our ansatz \eqref{HE-ansatz} and
\eqref{profile-ansatz} reflects the Feynman mechanism by connecting the
large-$t$ behavior of $F_1^q$ with the large-$x$ behavior of $q_v$.  The
stronger decrease of $F_1^d$ compared with $F_1^u$ seen in the data is
thus naturally explained by the faster decrease of $d_v$ compared with
$u_v$.  Correspondingly, our fit predicts a faster decrease of $-F_2^d$
compared with $F_2^u$ for $\sqrt{-t}$ above $1.5 \gev$.  It will be
interesting to see whether this is realized in Nature, once there are
precise data for $G_E^n$ at higher $-t$.

\section{Selected fit results}

\begin{figure}[hb]
\begin{center}
\includegraphics[width=0.46\textwidth]{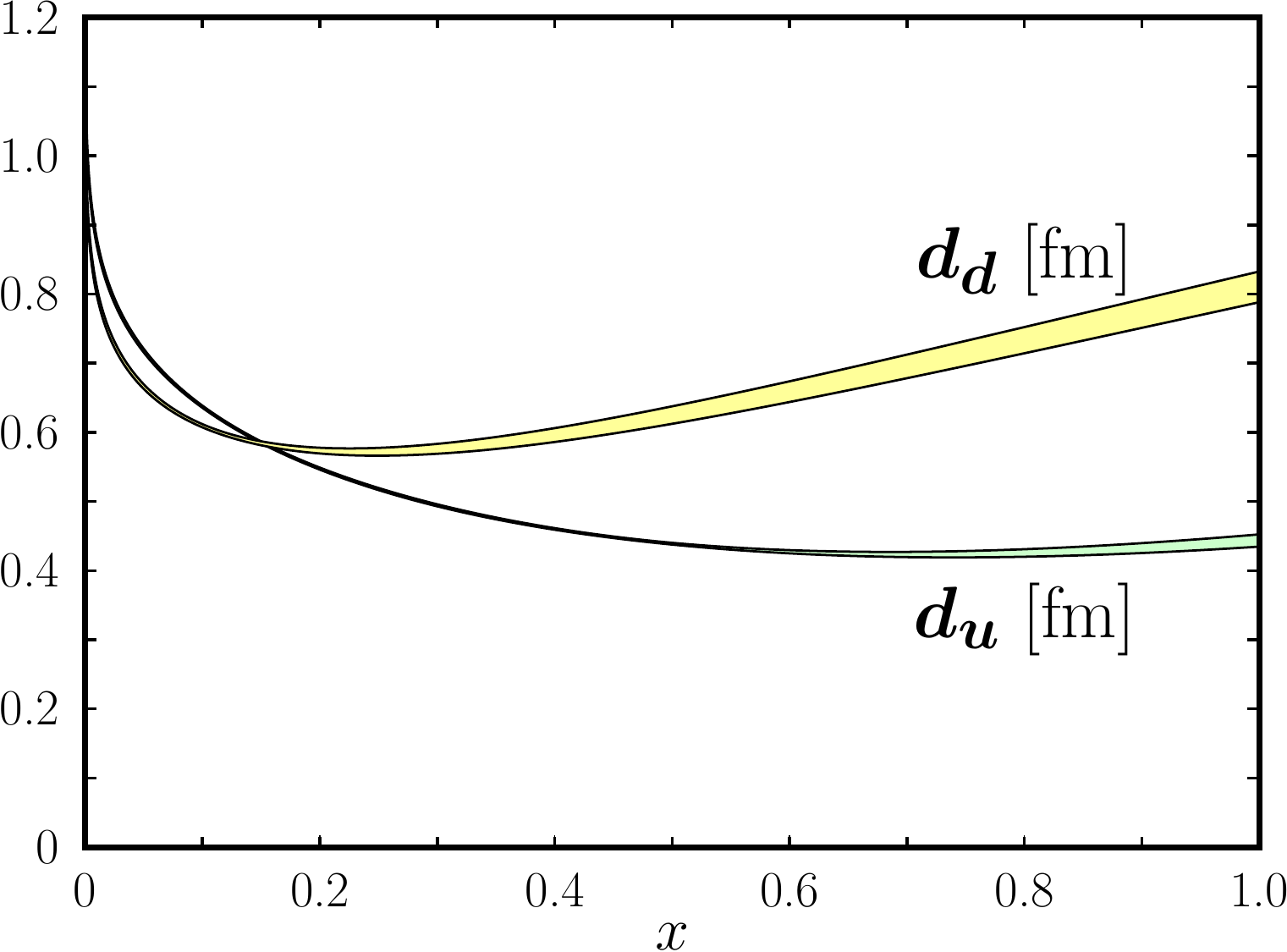}
\hspace{1em}
\includegraphics[width=0.46\textwidth]{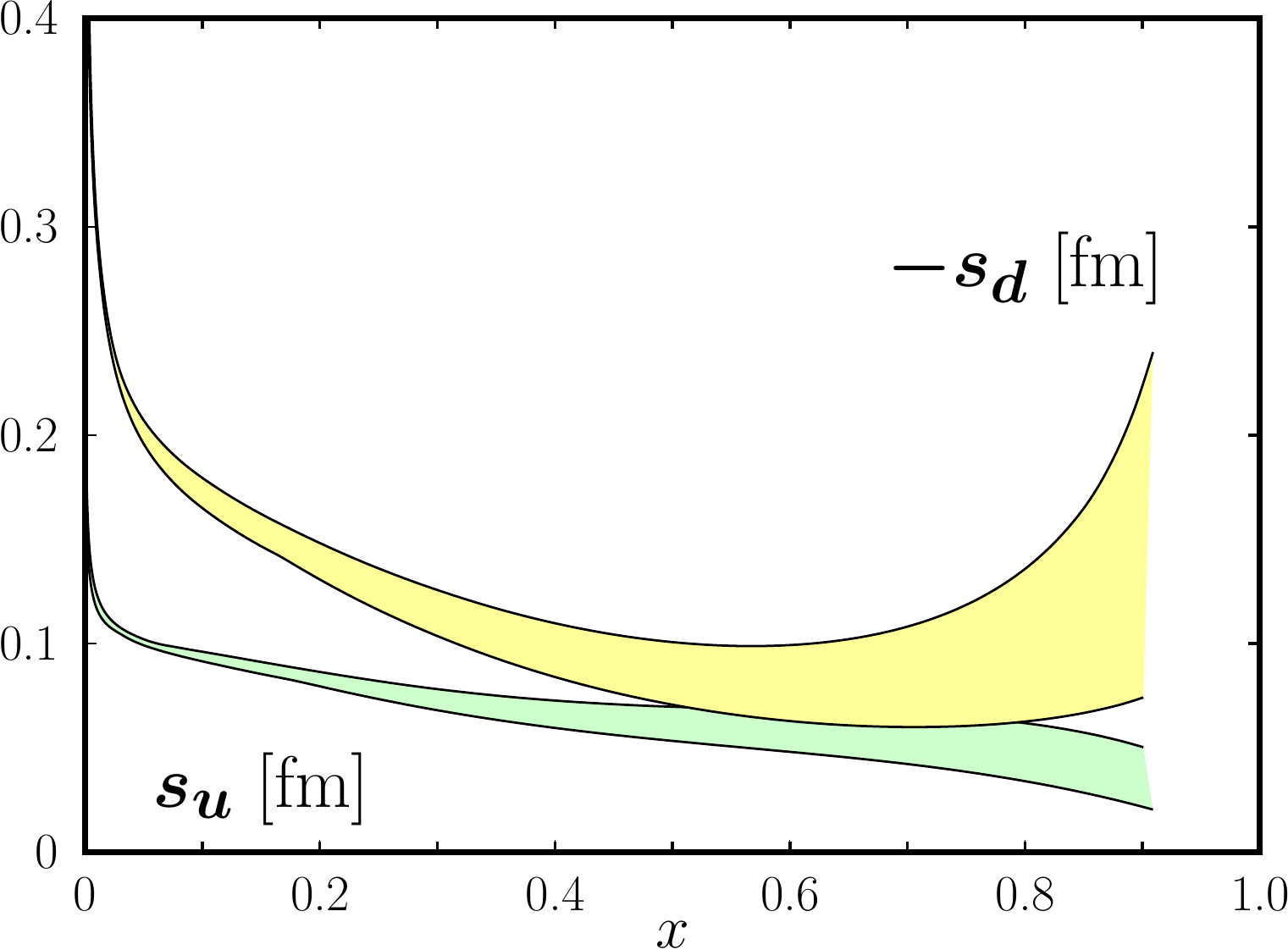}
\end{center}
\caption{\label{fig:distances} Results of our default GPD fit for the
  transverse distribution of valence quarks in the proton.  Left: average
  distance between the struck quark and the spectators.  Right: average
  shift of this distance in the $y$ direction when the proton is polarized
  in the positive $x$ direction.}
\end{figure}

Having extracted the GPDs $H_v^q$ and $E_v^q$, we can compute a variety of
quantities of interest for the structure of the proton.  The left panel in
figure~\ref{fig:distances} shows the average of the distance between the
struck parton and the spectators, defined by $d_q(x) = \bigl[ \langle
\vec{b}^2\rangle^q_x \bigr]{}_{}^{1/2} / (1-x)$, and the right panel shows
the average shift of this distance induced by transverse proton
polarization, $s_q(x) = \langle \vec{b}^y\rangle_x^q / (1-x)$.  A clear
difference between valence $d$ and $u$ quarks is seen for both quantities.

We can also evaluate the contribution of the total angular momentum
carried by quarks minus the corresponding contribution from antiquarks.
According to Ji's sum rule, this is given by
\begin{equation}
  \label{Ji-sum-rule}
J_v^q 
 = \frac{1}{2} \int_0^1 dx\; x \bigl[ q_v^{}(x) + e_v^q(x) \bigr] \,.
\end{equation}
From our GPD fit we obtain
\begin{align}
  \label{J-results}
J_v^u     &=  0.230^{ + 0.009}_{ - 0.024} \;,
&
J_v^d     &= -0.004^{ + 0.015}_{ - 0.016} \;,
&
J_v^{u+d} &=  0.226^{ + 0.004}_{ - 0.026} \;,
&
J_v^{u-d} &=  0.233^{ + 0.020}_{ - 0.038}
\end{align}
at the scale $\mu = 2\gev$, where the errors include the parametric
uncertainties of our default fit with the ABM 11 PDFs, as well as the
variation of results if we employ different PDFs or change other details
of the fit.\footnote{%
  In \protect\cite{Diehl:2013xca} the upper error on $J_v^d$ is
  erroneously quoted to be $+0.010$ instead of $+0.015$.}
These results are consistent with our previous work
\cite{Diehl:2004cx,Diehl:2005pk}, while using a much larger data set and
giving a more comprehensive estimate of the uncertainty.  They also agree
with a recent determination \cite{Bacchetta:2011gx} based on a
phenomenological extraction of the Sivers distributions and a model for
chromodynamic lensing, which gives $J_v^u = 0.214^{+0.009}_{-0.013}$ and
$J_v^d = -0.029^{+0.021}_{-0.008}$ if we add in quadrature the errors
quoted in \cite{Bacchetta:2011gx}.  We find it encouraging that two
model-dependent extractions using entirely different methods obtain
consistent results.
Subtracting from \eqref{J-results} the the helicity carried by valence
quarks as obtained from the DSSV set of polarized parton densities
\cite{deFlorian:2009vb}, we obtain orbital angular momentum contributions
\begin{align}
  \label{L-results}
L_v^u     &= -0.141^{ + 0.025}_{ - 0.033} \;,
&
L_v^d     &=  0.114^{ + 0.034}_{ - 0.035} \;,
&
L_v^{u+d} &= -0.027^{ + 0.029}_{ - 0.039} \;,
&
L_v^{u-d} &= -0.255^{ + 0.051}_{ - 0.061}
\end{align}
to the proton spin at $\mu = 2\gev$, where we have added in quadrature the
errors in \eqref{J-results} and those of the DSSV parameterization.  In
agreement with other analyses, we find a very small net orbital angular
momentum, which results from a cancellation between significant individual
contributions from $u$ and $d$ quarks.  We emphasize that the results
\eqref{J-results} and \eqref{L-results} are for the \emph{difference}
between quarks and antiquarks.  Using electromagnetic form factors, we
cannot access the contribution from antiquarks by itself.  Evaluating the
usual momentum sum rule for current PDF sets, we find that at $\mu =
2\gev$ the longitudinal momentum carried by sea quarks in the proton is of
similar size as the contribution from valence $d$ quarks.  If this
provides any guidance, then the contribution from sea quarks to Ji's sum
rule may not be negligible.


\section{Conclusions}

The electromagnetic nucleon form factors are among the best known
quantities describing proton structure.  With the experimental precision
that has been reached, several issues require further efforts before a
consistent and reliable picture can be established.  There are clear
inconsistencies between results for the ratio $G_E^p/G_M^p$, whose
resolution appears urgent to us.  The quantitative control of two-photon
exchange in the extraction of $G_M^p$, the measurement of $G_E^n$ or
$G_E^n/G_M^n$ at higher $-t$, as well as a quantitative determination of
the strangeness form factors, remain tasks for the future.

Using our selection of form factor data, we have determined a set of
interpolated form factor values up to $-t \simeq 3.4 \gev^2$, from which
one can compute various form factor combinations of interest.  We find
that the power-law ansatz \eqref{global-power-fit-fct} gives an excellent
and economical fit of all data.  A good description of the data can be
achieved with a fit of $u$ and $d$ valence quark GPDs, in which the
positivity conditions relating $E$ and $H$ play an essential role.  We
have used the resulting GPDs to quantify the transverse distribution of
valence quarks in the proton and to evaluate the total angular momentum
carried by valence quarks, along with several other applications presented
in \cite{Diehl:2013xca}.


\end{document}